\documentclass[sigconf]{acmart}

\usepackage{booktabs} 
\usepackage[skip=0pt]{caption}
\usepackage[most]{tcolorbox}
\usepackage{setspace}
\DeclareMathAlphabet{\pazocal}{OMS}{zplm}{m}{n}

\DeclareMathOperator \real{\mathbb{R}}

\newcommand{\Vs}{\pazocal{V}}

\newcommand{\Js}{\pazocal{J}}

\newcommand{\Cs}{\pazocal{C}}

\newcommand{\dev}{\textsc{dev}}

\newcommand{\ra}[1]{\renewcommand{\arraystretch}{#1}}

\usepackage{flowchart}
\usetikzlibrary{arrows}

\copyrightyear{2017}
\acmYear{2017}
\setcopyright{rightsretained}
\acmConference{SIGIR 2017 Workshop on Neural Information Retrieval (Neu-IR'17)}{}{August 7--11, 2017, Shinjuku, Tokyo, Japan} \acmPrice{}
\begin{document}
\title{Thread Reconstruction in Conversational Data using Neural Coherence Models}

\author{Dat Tien Nguyen}
\affiliation{%
  \institution{University of Amsterdam}
  \city{Amsterdam}
  \country{Netherlands}
}
\email{t.d.nguyen@uva.nl}

\author{Shafiq Joty}
\affiliation{%
  \institution{Nanyang Technological University}
  \city{Singapore}
  \country{Singapore}
}
\email{srjoty@ntu.edu.sg}

\author{Basma El Amel Boussaha}
\affiliation{%
  \institution{University of Nantes}
  \city{Nantes}
  \country{France}
  }
\email{basma.boussaha@univ-nantes.fr}

\author{Maarten de Rijke}
\affiliation{
  \institution{University of Amsterdam}
  \city{Amsterdam}
  \country{Netherlands}
  }
\email{derijke@uva.nl}

\renewcommand{\shortauthors}{}

\begin{abstract}

Discussion forums are an important source of information. They are often used to answer specific questions a user might have and to discover more about a topic of interest. Discussions in these forums may evolve in intricate ways, making it difficult for users to follow the flow of ideas. We propose a novel approach for automatically identifying the underlying thread structure of a forum discussion. Our approach is based on a neural model that computes coherence scores of possible reconstructions and then selects the highest scoring, i.e., the most coherent one. Preliminary experiments demonstrate promising results outperforming a number of strong baseline methods.

\end{abstract}

%
%

\keywords{Thread reconstruction; Coherence model; Convolutional neural network}


\maketitle

\section{Introduction}
Discussion forums are an important source of information, both to help a user answer specific information needs they might have and to help users who want to explore a topic without having a specific question in mind. Discussions in these forums are typically composed of multiple inter-woven threads, regardless of whether that threaded structure is made explicit in the representation and presentation of the conversational data~\cite{Wang:2011:PTD}. Recovering the underlying thread structure is helpful for users of discussion forums as it makes possible to disentangle discussions related to subtopics and/or to particular conversational goals~\cite{wang-recovering-2008}.

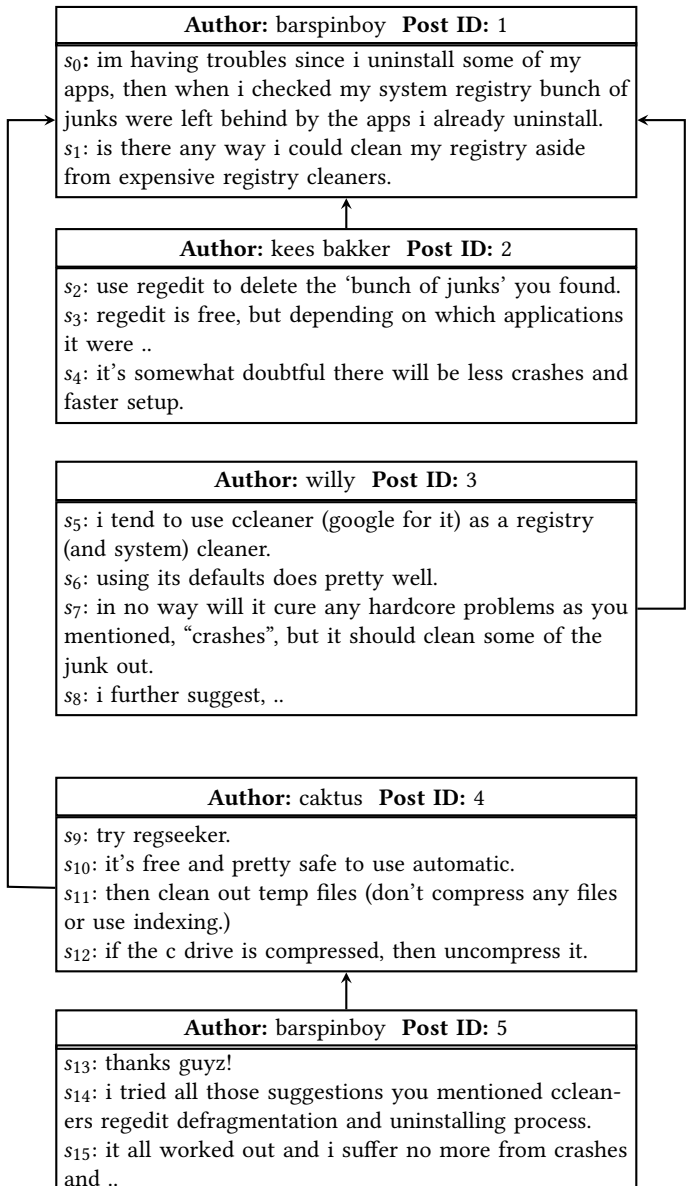
\begin{figure}
\begin{tikzpicture}[>=stealth, thick]

\node (A1) at (0,1.25) [draw, process, text width=7.5cm, minimum height=0.5cm, align=center] 
{\textbf{Author:} barspinboy \textbf{ Post ID:} 1};

\node (A2) at (0,0) [draw, process, text width=7.5cm, minimum height=0.5cm, align=left] 
{ \textbf{$s_0$:} im having troubles since i uninstall some of my apps, then when i checked my system registry bunch of junks were left behind by the apps i already uninstall.\\ 
$s_1$: is there any way i could clean my registry aside from expensive registry cleaners. };

\node (B1) at (0,-1.7) [draw, process, text width=7.5cm, minimum height=0.5cm, align=center] 
{\textbf{Author:} kees bakker \textbf{ Post ID:} 2};

\node (B2) at (0,-3.0) [draw, process, text width=7.5cm, minimum height=0.5cm, align=left] 
{
$s_2$: use regedit to delete the `bunch of junks' you found.\\ 
$s_3$: regedit is free, but depending on which applications it were .. \\
$s_4$: it's somewhat doubtful there will be less crashes and faster setup.};

\node (C1) at (0,-4.8) [draw, process, text width=7.5cm, minimum height=0.5cm, align=center] 
{\textbf{Author:} willy \textbf{ Post ID:} 3};

\node (C2) at (0,-6.5) [draw, process, text width=7.5cm, minimum height=0.5cm, align=left] 
{$s_5$: i tend to use ccleaner (google for it) as a registry (and system) cleaner. \\ 
$s_6$: using its defaults does pretty well.  \\
$s_7$: in no way will it cure any hardcore problems as you mentioned, ``crashes'', but it should clean some of the junk out.  \\
$s_8$: i further suggest, .. };

\node (D1) at (0,-9) [draw, process, text width=7.5cm, minimum height=0.5cm, align=center] 
{\textbf{Author:} caktus \textbf{ Post ID:} 4};

\node (D2) at (0,-10.3) [draw, process, text width=7.5cm, minimum height=0.5cm, align=left] 
{ $s_{9}$: try regseeker. \\ 
$s_{10}$: it's free and pretty safe to use automatic.  \\
$s_{11}$: then clean out temp files (don't compress any files or use indexing.)  \\
$s_{12}$: if the c drive is compressed, then uncompress it.};

\node (E1) at (0,-12.1) [draw, process, text width=7.5cm, minimum height=0.5cm, align=center] 
{\textbf{Author:} barspinboy \textbf{ Post ID:} 5 };

\node (E2) at (0,-13.3) [draw, process, text width=7.5cm, minimum height=0.5cm, align=left ] 
{$s_{13}$: thanks guyz! \\
$s_{14}$: i tried all those suggestions you mentioned ccleaners regedit defragmentation and uninstalling process. \\
$s_{15}$: it all worked out and i suffer no more from crashes and ..};

\coordinate (x) at (-4.5,-10.2);
\coordinate (y) at (-4.5,0);

\coordinate (z) at (4.5,0);
\coordinate (k) at (4.5,-6.5);

\draw[->] (C2) -- (k) -- (z) |- (A2);

\draw[->] (B1) -- (A2);
\draw[->] (E1) -- (D2);

\draw[->] (D2) -- (x) -- (y) |- (A2);

\end{tikzpicture}
\caption{A truncated forum thread from CNET with five comments by temporal order. \emph{Reply-to} links between posts are denoted by arrowed edges.}
\label{fig:example}
\end{figure}

More precisely, the thread reconstruction task is defined as follows. Given a discussion stream in which messages are sorted by posting time, the thread reconstruction task is to construct the thread tree. The edges in the tread tree identify which of the previously contributed posts the current post replies to. For example, consider the example thread from CNET forum site\footnote{https://www.cnet.com/} in Figure \ref{fig:example}, where we have five ($p$)osts. The thread has a tree structure with three branches: $p_1 \leftarrow p_2$, $p_1 \leftarrow p_3$ and $p_1 \leftarrow p_4 \leftarrow p_5$. Given the posts $\{p_1, p_2, \ldots, p_5\}$, our goal in thread reconstruction is to recover the underlying tree structure of the thread. 

Several methods have been proposed for the thread reconstruction task \cite{yi:2008,Wang:2011:PTD,wang-recovering-2008}. These methods learn an edge-level classifier to decide for a possible connection using features like distance in position/time, cosine similarity between comments, etc. However, these models suffer from the limitation that they consider one edge at a time rather than the global tree structure of the thread. Modeling edges locally disregards  interactions between all possible edges and can lead to suboptimal solutions. In contrast, in this paper we propose to model an entire thread for the reconstruction task. We propose to use a neural coherence model \cite{dat-joty:2017} from natural language processing (NLP) for scoring candidate tree hypotheses.

Coherence models \cite{Barzilay:2008,Elsner:2011} were originally proposed for coherence assessment of monologues (e.g., news articles, books). However, forum conversations are different from monologues in the sense that information flow in these conversations are not sequential; topics in these conversations are often interleaved in the temporal order of the comments \cite{Joty:2013,louis2015conversation}. For example, in the thread in Figure \ref{fig:example}, there are three possible subconversations each corresponding to a branch. The branch $p_1 \leftarrow p_2$ suggests \emph{using regedit}, the branch $p_1 \leftarrow p_3$ suggests \emph{ccleaner}, and the third branch suggests \emph{regseeker}.  Beacause of these differences, when applied directly to these conversations, the coherence models may not perform as we expect. Furthermore, these models are not specifically trained for the reconstruction task.   

In this paper, we make the following contributions. First, we hypothesize that coherence models should consider the thread structure of a conversation and we extend the original grid representation proposed by Barzilay and Lapata \cite{Barzilay:2008} to encode the thread structure of a forum conversation. Then we train a convolutional neural network (CNN) model with pairwise ranking using the grid representation for the thread reconstruction task. Our method considers the whole thread structure at once, and computes coherence scores for all possible candidate trees. The highest scoring tree corresponds to the predicted tree structure for the given thread. 

We evaluated our approach on discussion threads from CNET. The results show that our method is quite promising outperforming several strong baselines on this dataset.

\section{Coherence Models}

In this section, we give a brief overview of the coherence models that were originally proposed for monologues (e.g., news articles) and that are related to our work. In the next section, we propose extensions to these models for forum-like conversations that we use for thread reconstruction.

\subsection{Entity Grid and Its Extensions}

Barzilay and Lapata \cite{Barzilay:2008} proposed an entity-based model for representing and assessing text coherence. Their model represents a text by a matrix called \emph{entity grid} that captures transitions of  entities (i.e., noun phrases) across sentences. As shown in Table \ref{table:grid}, the rows of the grid correspond to sentences, and the columns correspond to entities appearing in the text. Each entry $G_{i,j}$ in the entity grid represents the syntactic role that entity $e_j$ plays in sentence $s_i$, which can be one of: subject (\textbf{S}), object (\textbf{O}), or other (\textbf{X}). Entities not appearing in a sentence are marked by a special symbol (\textbf{-}). 

To represent the grid using a feature vector, Barzilay and Lapata \cite{Barzilay:2008} compute probability for each local entity transition of length $k$ (i.e., $\{S,O,X,-\}^k$), and represent each grid by a vector of $4^k$ transitions probabilities. 
Coherence assessment is then formulated as a ranking task in an SVM preference ranking framework \cite{Joachims:2002}.

A number of extensions of the basic entity grid model have been proposed. Elsner and Charniak \cite{Elsner:2011} extended the basic grid to distinguish between entities of different types by incorporating entity-specific features like named entity, noun class, modifiers, etc. Feng and Hirst \cite{Feng:2012} used the basic grid representation, but improved its learning to rank scheme. Their model learns not only from original document and its permutations but also from ranking preferences among the permutations themselves.    

\subsection{Neural Entity Grid Model} \label{subsec:cnn-egrid}

Although the entity grid and its extensions have been successfully applied to many downstream applications including coherence rating \cite{Barzilay:2008}, readability assessment \cite{Pitler:2010,Barzilay:2008}, essay scoring \cite{Burstein:2010}, and story generation \cite{McIntyre:2010}, they have some limitations. First, they use discrete representation for grammatical roles and features, which leads to the so-called \textbf{curse of dimensionality} problem \cite{Bengio03}. In particular, to model transitions of length $k$ with $\Cs$ different grammatical roles, the basic entity grid model needs to compute $\Cs^k$ transition probabilities from a \emph{single} grid. The estimated distribution becomes sparse as $k$ increases, which  prevents the model from considering longer transitions -- existing models typically use $k \le 3$. Second, these models compute feature representations from entity grids in a task-agnostic way. Decoupling feature extraction from the target task can limit the model's capacity to learn task-specific features. 

\begin{figure*}[t!]
\centering
\includegraphics[width=0.8\textwidth]{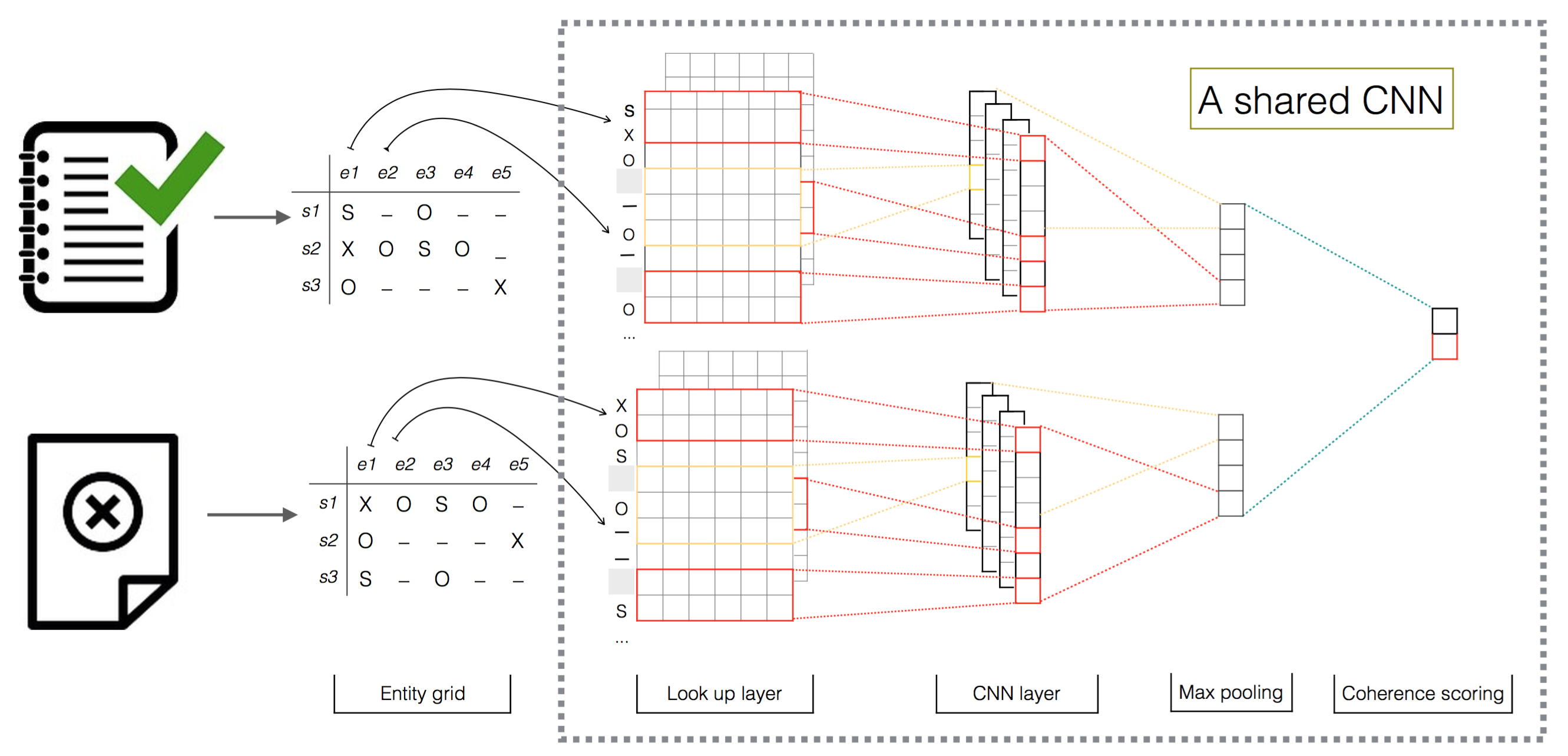}
\caption{Neural model for coherence scoring and the pairwise training method [taken from our previous work \cite{dat-joty:2017}].}
\label{fig:cnn_model}
\end{figure*}

To deal with the above issues of entity grid models, we \cite{dat-joty:2017} recently proposed a neural extension to the grid models. As shown in Figure \ref{fig:cnn_model}, the neural model takes an extracted entity grid as input, and transforms each grammatical role in the grid into a distributed representation by looking up a shared embedding matrix $E$ $\in$ $\real^{|\Vs| \times d}$, where $\Vs = \{S,O,X,-\}$ is a set of grammatical roles, and $d$ is the embedding dimensions. The embedding vectors produced by the lookup layer are combined by subsequent layers of the network to generate a coherence score for the document. The network uses a \emph{convolutional} layer, which applies $N$ \emph{filters} to get $N$ different \emph{feature maps}. The abstract features in each feature map are then pooled using a \emph{max-pooling} operation. The pooled features are then used for coherence scoring at the final layer of the model. 


Convolution learns to compose local transitions of a grid into higher-level representations, while max-pooling captures the most salient local features from each feature map. Since the convolution-pooling operates over the distributed representation of grid entries, compared to traditional grid models, the transition length $k$ can be sufficiently large to capture long-range dependencies without overfitting on the training data. Also, the embedding vectors and the convolutional filters are learned from all training documents as opposed to a single document in traditional grid models, which helps the neural model to obtain better  generalization and robustness. The evaluation on three different coherence assessment tasks demonstrates the superiority of the neural model yielding state of the art results. In this work, we therefore extend the neural model for forum-like conversations and use it for thread reconstruction.

\section{Neural Coherence Model for Forum Threads}

The main difference between forum conversations and monologues is that the information flow in forum conversations is often not sequential as in monologue. As a result, the coherence models that are originally developed for monologues may not perform as expected when they are directly applied to threaded conversations \cite{Joty:2013}. We hypothesize that the coherence models should consider the conversational structure in the form of ``reply-to'' relations between comments as shown by a tree structure in Figure \ref{fig:example}. In the following subsections, we describe how we extend the neural entity grid model to incorporate the tree structure of a thread.

\subsection{Entity Grid for Forum Threads}

The thread structure in Figure \ref{fig:example} has a tree structure, where nodes represent comments and edges represent ``reply-to'' links between comments. Since entity grid models operate at the sentence level, we construct the conversational thread at the sentence level. We do this by linking the boundary sentences across comments and by linking sentences in the same comment chronologically; i.e., we connect the first sentence of comment $c_j$ to the last sentence of comment $c_i$ if $c_j$ is a reply to $c_i$, and sentence $s_{t+1}$ is linked to $s_t$ if both $s_{t}$ and $s_{t+1}$ are in the same comment.

To encode a sentence-level conversation tree into an entity grid, we propose couple of modifications to the original entity grid representation. In the modified representation as shown in Table \ref{table:grid}, rows represent depth levels of the conversation tree as opposed to sentences in the original grid. An entry $G_{i,j}$ in our conversational entity grid represents the sequence of grammatical roles (left to right) that the entity $e_j$ plays in the sentences occurring at the $j$-th level of the conversation tree. For instance, our example tree has three sentences $s_3$, $s_6$ and $s_{10}$ at depth level $3$. The entity \emph{REGEDIT} has the role of a \emph{Subject}, a \emph{not present} and a \emph{not present}, respectively, in these three sentences, thus encoded as `\emph{S{-}{-}}' in the entity grid.

\begin{table*}[htb]
\caption{Transition of some entities across tree structure of the thread example. Legend: S stands for subject, O for object, X for a role other than subject or object, and -- means that an entity does not appear in the sentence.}
\label{table:grid}
\centering
\begin{tabular}{cccccccccccccccccccc}
 \multicolumn{3}{c}{Tree structure} & depth& {\rotatebox{90}{CLEANER}} & {\rotatebox{90}{REGEDIT}} & {\rotatebox{90}{TROUBLES}} & {\rotatebox{90}{SYSTEM}} & {\rotatebox{90}{JUNKS}} & {\rotatebox{90}{APPS}} & {\rotatebox{90}{REGISTRY}} & {\rotatebox{90}{BUNCH}} \\ 
\toprule
 \multicolumn{3}{c}{ $s_{0}$} & 0 & $-$    &$-$&$-$ &O& X&X&$O$&$O$ \\ 
\multicolumn{3}{c}{ $s_{1}$} & 1 & O       &$-$&$-$&$-$&$-$&$-$&$O$&$-$ \\ 
$s_{2}$ & $s_{5}$ & $s_{9}$ &  2  & $-$O$-$ &O$--$ &$---$& $---$& X$--$& $---$&$-O-$&O$--$ \\ 
$s_{3}$ & $s_{6}$ & $s_{10}$ & 3 & $---$   &S$--$ &$---$& $---$& $---$& $---$ &$---$&$---$ \\ 
$s_{4}$ & $s_{7}$ & $s_{11}$ & 4 & $---$   &$---$ &$---$& $---$& $-X-$& $---$ &$---$&$---$ \\ 
& $s_{8}$& $s_{12}$ & 5          & $--$    &$--$ &$--$&  $--$& $--$&  $--$  &$--$&$--$ \\
\bottomrule
\end{tabular}
\end{table*}

\subsection{Thread Reconstruction}
\label{subsection:threadreconstruction}
The conversational entity grid captures transition of entities in terms of their grammatical roles in a conversation tree. We believe this representation can be quite useful for thread reconstruction -- i.e., discovering the latent structure of a forum thread.

We train a convolutional neural network (we refer to our model as {\bf Grid-CNN} for the rest of this paper) using the conversational entity grid representation for the thread reconstruction task. The CNN model has the same structure as described in Section \ref{subsec:cnn-egrid}. 

We use a \emph{pairwise ranking} approach \cite{collobert2011natural} to train the Grid-CNN model. For a given number of comments in a gold tree, we first construct a set of valid candidate trees. A valid tree is one that respects the chronological order of the comments in a thread -- for example, a comment can only reply to a comment that comes before in the temporal order. The training set comprises \emph{ordered} pairs $(T_i, T_j)$, where thread $T_i$ is a \emph{true (gold)} tree and $T_j$ is a valid but \emph{false} tree. We seek to find model parameters that assign a higher score to $T_i$ than to $T_j$. We minimize the following ranking loss:
\begin{equation}
\Js(\theta)=  \max \{0, 1 - \phi(G_i|\theta) + \phi(G_j|\theta)\} \label{loss}
\end{equation}
where $G_i$ and $G_j$ are the conversational entity grids corresponding to threads $T_i$ and $T_j$, respectively, and $\theta$ defines the model parameters including the embedding matrix and the weight vectors. 

During testing, our Grid-CNN model predicts coherence scores for all the possible candidate trees given the posts in a thread, and the tree with the highest score is considered to be the underlying structure of the thread.




\section{Experiment}

Our main research question is to determine whether text coherence can benefit thread reconstruction; we want to assess at two levels, the level of trees (``did we reconstruct the discourse correctly'') and the level of edges (``did we identify individual replies correctly'').

\subsection{Experimental setup}

As a preliminary experiment, we collect 2,200 threads from the CNET dataset made available in previous research~\cite{kim-tagging-2010} that have less than 6 posts. 
We remove from the raw data some meta 
information and keep the temporal order (post ID) of the dataset.
We leave 1,500 threads for training, 200 threads as our development set and the rest for testing. See Table~\ref{table:corpora} for further details about the corpus.

\begin{table}[htb]
\caption{Corpus statistics. \emph{Non-trivial} replies are posts that reply to other posts except the first post.}
\label{table:corpora}
\begin{tabular}{cccc}
\toprule
 \# Threads & Avg. \# Posts & Avg. \# Sent & Non-trivial replies  \\ 
\midrule
 2,200  & 3.6 & 27.64 & 57\%\\ 
\bottomrule
\end{tabular}
\end{table}


We train the Grid-CNN model by optimizing the pairwise ranking loss in Equation~\ref{loss} %
using the gradient-based online learning algorithm RMSprop\cite{Tieleman12}.
\footnote{Other adaptive algorithms, e.g., ADAM \cite{KingmaB14}, ADADELTA \cite{Zeiler12} gave similar results.}
We use up to 25 epochs. We use dropout \cite{Srivastava14a} of hidden units to avoid overfitting, and do \emph{early stopping} by observing accuracy on the \dev\ set -- if the performance (accuracy score) does not increase for $10$ consecutive epochs, we stop training and pick the best model recorded so far. We search for optimal mini\textbf{batch} size in $\{16, 32, 64, 128\}$, \textbf{emb}edding size in $\{50, 80, 100, 200\}$, \textbf{dropout} rate in $\{0.2, 0.3, 0.5\}$, \textbf{filter} number in $\{100, 150, 200, 300\}$, \textbf{win}dow size in $\{3,4, 5, 6, 7, 8\}$, and \textbf{pool}ing length in $\{4, 5, 6, 7\}$. The best model (see Table ~\ref{tab:best-cnn-setting}) on the development set is then used for the final evaluation on the test set.

\begin{table}[t!]
\ra{1.2}
\resizebox{1.0\linewidth}{!}{
\begin{tabular}{l|cccccc}
\toprule
& Batch & Emb.  & Dropout &  Filter & Win.    & Pool \\ 
\midrule 
Grid-CNN  & 64 & 100 & 0.5 & 150 & 6 & 6 \\
\bottomrule 
\end{tabular}}
\caption{Optimal hyper-parameter setting for our neural models based on development set accuracy.}
\label{tab:best-cnn-setting}
\end{table}

For comparison, we use a number of simple but well performing baselines: 

\begin{description}
\item[{\bf All-previous}] Create thread structure by linking a comment to its previous comment.
\item[{\bf All-first}] Create thread structure by linking all the comments to the first comment.
\item[{\bf COS-sim}] Create thread structure by linking a comment to one of the previous comments with which it has the highest cosine similarity.
\end{description}


We consider two variations of the thread reconstruction task: tree-level and edge-level. For the tree-level version of the thread reconstruction task, the model predicted tree should match entirely the original structure of the thread. For the edge-level version, we measure the post-level (the link between two posts) classification performance. If a link appears in the predicted tree and the original one, we count as a true prediction; otherwise it is a false case.

\subsection{Results and discussion}

Table~\ref{tab:recon_results} presents the results for the baselines and our method, Grid-CNN. The {\bf COS-sim} gets the lowest score on both reconstruction tasks. This means reply structure between two posts does not rely only on their term matching patterns. The other two baselines {\bf All-previous} and {\bf All-first} perform quite well on our dataset. This is not surprising since it is very common in this kind of forums that most participants reply either to the previous post or to the first post that asks the question. 

Our neural method gets promising results,  yielding substantial improvements over the baselines in all cases. The {\bf Grid-CNN} model delivers relative improvements from 32\% to 57\% in accuracy for the tree-level reconstruction task. It also outperforms the baselines in the edge-level prediction task with improvements from 4\% to 13\% in $F_1$-score and from 1\% to 12\% in accuracy.

We further manually inspected the false prediction cases for our method. We observed that most of the false trees fall into the trivial structures (All-previous or All-first). This could be due to the dominance of these cases in our training data -- 40.07\% of the posts reply to the first post and 76.29\% reply to the previous post.

\begin{table}[htb]
\centering
\caption{Performance on the thread reconstruction task.} 
\label{tab:recon_results}
\begin{tabular}{lcll}
\toprule
 & \textbf{Tree-level} &  \multicolumn{2}{c}{\textbf{Edge-level}} \\
\cmidrule(lr){2-2}\cmidrule(lr){3-4}
 & Acc & $F_1$ & ~Acc \\  
\midrule
All-previous  & 20.00 & 58.45 & 65.62\\
All-first  & 17.60 & 54.90 &60.27\\
COS-sim  & 16.80 & 53.58 & 58.75  \\
\midrule
Grid-CNN   & \bf 26.40 & \bf 60.55 & \bf 66.12 \\
\bottomrule
\end{tabular}
\end{table}


\section{Related Work}

Several previous studies treat thread reconstruction as an edge-level classification problem. \citet{yi:2008} use cosine similarity between posts and exploit temporal order information (e.g., time distance, post distance) to recover the thread structure. 
\citet{erik:conf} consider thread reconstruction as a classification problem.
They train a decision tree classifier based on some basic features such as reply distance in number of posts, time distance, cosine similarity and thread lengths, etc. Their model takes a pair of posts as input and predicts the link between them. A joint model using dependency parsing and conditional random fields was proposed to predict links between two posts and their dialogue acts \cite{Wang:2011:PTD}. \citet{Dehghani:2013} works on reconstructing tree structure of conversation threads in email data.


In contrast to previous approaches, we treat thread reconstruction as a ranking problem and use a neural coherence model to rank  all possible candidate trees. We show that modeling coherence of threaded conversations is an effective approach to thread reconstruction.

\section{Conclusions}
This paper introduces a novel approach to solve thread reconstruction problem in discussion forums. Our method uses a neural coherence model based on an entity grid representation and a convolutional neural network (CNN).  
First, we extend the original grid representation to encode the thread structure of a forum conversation. Then we train a CNN model with pairwise ranking using the grid representation for the thread reconstruction task. Our method considers the whole thread structure at once, and computes coherence scores for all possible candidate trees. The highest scoring tree is returned as the predicted tree structure.  




We evaluated our approach on discussion threads from CNET forum site. The result shows that our method is very promising. It significantly improves performance over trivial baselines, particularly for the tree-level accuracy. In the future, we would like to experiment with larger datasets containing threads with many posts. We also plan to integrate other discourse structures like dialogue acts into our model to get further improvements.

\begin{spacing}{1}
\medskip\noindent\small
\textbf{Acknowledgments.}
This research was supported by
Ahold Delhaize,
Amsterdam Data Science,
the Bloomberg Research Grant program,
the Criteo Faculty Research Award program,
the Dutch national program COMMIT,
Elsevier,
the European Community's Seventh Framework Programme (FP7/2007-2013) under
grant agreement nr 312827 (VOX-Pol),
the Microsoft Research Ph.D.\ program,
the Netherlands Institute for Sound and Vision,
the Netherlands Organisation for Scientific Research (NWO)
under pro\-ject nrs
612.\-001.\-116, 
HOR-11-10, 
CI-14-25, 
652.\-002.\-001, 
612.\-001.\-551, 
652.\-001.\-003, 
and
Yandex.
All content represents the opinion of the authors, which is not necessarily shared or endorsed by their respective employers and/or sponsors.
\end{spacing}

\bibliographystyle{ACM-Reference-Format}
\bibliography{coherence} 


\begin{thebibliography}{00}


\ifx \showCODEN    \undefined \def \showCODEN     #1{\unskip}     \fi
\ifx \showDOI      \undefined \def \showDOI       #1{#1}\fi
\ifx \showISBNx    \undefined \def \showISBNx     #1{\unskip}     \fi
\ifx \showISBNxiii \undefined \def \showISBNxiii  #1{\unskip}     \fi
\ifx \showISSN     \undefined \def \showISSN      #1{\unskip}     \fi
\ifx \showLCCN     \undefined \def \showLCCN      #1{\unskip}     \fi
\ifx \shownote     \undefined \def \shownote      #1{#1}          \fi
\ifx \showarticletitle \undefined \def \showarticletitle #1{#1}   \fi
\ifx \showURL      \undefined \def \showURL       {\relax}        \fi
\providecommand\bibfield[2]{#2}
\providecommand\bibinfo[2]{#2}
\providecommand\natexlab[1]{#1}
\providecommand\showeprint[2][]{arXiv:#2}

\bibitem[\protect\citeauthoryear{Aumayr, Jeffrey, and Hayes}{Aumayr
  et~al\mbox{.}}{2011}]%
        {erik:conf}
\bibfield{author}{\bibinfo{person}{Erik Aumayr}, \bibinfo{person}{Chan
  Jeffrey}, {and} \bibinfo{person}{Conor Hayes}.}
  \bibinfo{year}{2011}\natexlab{}.
\newblock \showarticletitle{Reconstruction of Threaded Conversations in Online
  Discussion Forums}. In \bibinfo{booktitle}{{\em Proceedings of the Eleventh
  International Conference on Web and Social Media, {ICWSM} 2011.}}
\newblock


\bibitem[\protect\citeauthoryear{Barzilay and Lapata}{Barzilay and
  Lapata}{2008}]%
        {Barzilay:2008}
\bibfield{author}{\bibinfo{person}{Regina Barzilay} {and}
  \bibinfo{person}{Mirella Lapata}.} \bibinfo{year}{2008}\natexlab{}.
\newblock \showarticletitle{Modeling Local Coherence: An Entity-Based
  Approach}.
\newblock \bibinfo{journal}{{\em Computational Linguistics\/}}
  \bibinfo{volume}{34}, \bibinfo{number}{1} (\bibinfo{year}{2008}),
  \bibinfo{pages}{1--34}.
\newblock
\showURL{%
\url{http://www.aclweb.org/anthology/J08-1001}}


\bibitem[\protect\citeauthoryear{Bengio, Ducharme, Vincent, and Janvin}{Bengio
  et~al\mbox{.}}{2003}]%
        {Bengio03}
\bibfield{author}{\bibinfo{person}{Yoshua Bengio}, \bibinfo{person}{R{\'e}jean
  Ducharme}, \bibinfo{person}{Pascal Vincent}, {and} \bibinfo{person}{Christian
  Janvin}.} \bibinfo{year}{2003}\natexlab{}.
\newblock \showarticletitle{A Neural Probabilistic Language Model}.
\newblock \bibinfo{journal}{{\em J. Mach. Learn. Res.\/}}  \bibinfo{volume}{3}
  (\bibinfo{date}{March} \bibinfo{year}{2003}).
\newblock
\showISSN{1532-4435}
\showURL{%
\url{http://dl.acm.org/citation.cfm?id=944919.944966}}


\bibitem[\protect\citeauthoryear{Burstein, Tetreault, and Andreyev}{Burstein
  et~al\mbox{.}}{2010}]%
        {Burstein:2010}
\bibfield{author}{\bibinfo{person}{Jill Burstein}, \bibinfo{person}{Joel
  Tetreault}, {and} \bibinfo{person}{Slava Andreyev}.}
  \bibinfo{year}{2010}\natexlab{}.
\newblock \showarticletitle{Using Entity-based Features to Model Coherence in
  Student Essays}. In \bibinfo{booktitle}{{\em Human Language Technologies: The
  2010 Annual Conference of the North American Chapter of the Association for
  Computational Linguistics}} {\em (\bibinfo{series}{HLT '10})}.
  \bibinfo{publisher}{Association for Computational Linguistics},
  \bibinfo{address}{Los Angeles, California}, \bibinfo{pages}{681--684}.
\newblock


\bibitem[\protect\citeauthoryear{Collobert, Weston, Bottou, Karlen,
  Kavukcuoglu, and Kuksa}{Collobert et~al\mbox{.}}{2011}]%
        {collobert2011natural}
\bibfield{author}{\bibinfo{person}{Ronan Collobert}, \bibinfo{person}{Jason
  Weston}, \bibinfo{person}{L{\'e}on Bottou}, \bibinfo{person}{Michael Karlen},
  \bibinfo{person}{Koray Kavukcuoglu}, {and} \bibinfo{person}{Pavel Kuksa}.}
  \bibinfo{year}{2011}\natexlab{}.
\newblock \showarticletitle{Natural language processing (almost) from scratch}.
\newblock \bibinfo{journal}{{\em The Journal of Machine Learning Research\/}}
  \bibinfo{volume}{12} (\bibinfo{year}{2011}), \bibinfo{pages}{2493--2537}.
\newblock


\bibitem[\protect\citeauthoryear{Dehghani, Shakery, Asadpour, and
  Koushkestani}{Dehghani et~al\mbox{.}}{2013}]%
        {Dehghani:2013}
\bibfield{author}{\bibinfo{person}{Mostafa Dehghani}, \bibinfo{person}{Azadeh
  Shakery}, \bibinfo{person}{Masoud Asadpour}, {and} \bibinfo{person}{Arash
  Koushkestani}.} \bibinfo{year}{2013}\natexlab{}.
\newblock \showarticletitle{A Learning Approach for Email Conversation Thread
  Reconstruction}.
\newblock \bibinfo{journal}{{\em J. Inf. Sci.\/}} \bibinfo{volume}{39},
  \bibinfo{number}{6} (\bibinfo{date}{Dec.} \bibinfo{year}{2013}),
  \bibinfo{pages}{846--863}.
\newblock
\showISSN{0165-5515}
\showDOI{%
\url{https://doi.org/10.1177/0165551513494638}}


\bibitem[\protect\citeauthoryear{Elsner and Charniak}{Elsner and
  Charniak}{2011}]%
        {Elsner:2011}
\bibfield{author}{\bibinfo{person}{Micha Elsner} {and} \bibinfo{person}{Eugene
  Charniak}.} \bibinfo{year}{2011}\natexlab{}.
\newblock \showarticletitle{Extending the Entity Grid with Entity-specific
  Features}. In \bibinfo{booktitle}{{\em Proceedings of the 49th Annual Meeting
  of the Association for Computational Linguistics: Human Language
  Technologies: Short Papers - Volume 2}} {\em (\bibinfo{series}{HLT '11})}.
  \bibinfo{publisher}{Association for Computational Linguistics},
  \bibinfo{address}{Portland, Oregon}, \bibinfo{pages}{125--129}.
\newblock


\bibitem[\protect\citeauthoryear{Feng and Hirst}{Feng and Hirst}{2012}]%
        {Feng:2012}
\bibfield{author}{\bibinfo{person}{Vanessa~Wei Feng} {and}
  \bibinfo{person}{Graeme Hirst}.} \bibinfo{year}{2012}\natexlab{}.
\newblock \showarticletitle{Extending the Entity-based Coherence Model with
  Multiple Ranks}. In \bibinfo{booktitle}{{\em Proceedings of the 13th
  Conference of the European Chapter of the Association for Computational
  Linguistics}} {\em (\bibinfo{series}{EACL '12})}.
  \bibinfo{publisher}{Association for Computational Linguistics},
  \bibinfo{address}{Avignon, France}, \bibinfo{pages}{315--324}.
\newblock


\bibitem[\protect\citeauthoryear{Joachims}{Joachims}{2002}]%
        {Joachims:2002}
\bibfield{author}{\bibinfo{person}{Thorsten Joachims}.}
  \bibinfo{year}{2002}\natexlab{}.
\newblock \showarticletitle{Optimizing Search Engines Using Clickthrough Data}.
  In \bibinfo{booktitle}{{\em Proceedings of the Eighth ACM SIGKDD
  International Conference on Knowledge Discovery and Data Mining}} {\em
  (\bibinfo{series}{KDD '02})}. \bibinfo{publisher}{ACM},
  \bibinfo{address}{Edmonton, Alberta, Canada}, \bibinfo{pages}{133--142}.
\newblock
\showISBNx{1-58113-567-X}


\bibitem[\protect\citeauthoryear{Joty, Carenini, and Ng}{Joty
  et~al\mbox{.}}{2013}]%
        {Joty:2013}
\bibfield{author}{\bibinfo{person}{Shafiq Joty}, \bibinfo{person}{Giuseppe
  Carenini}, {and} \bibinfo{person}{Raymond~T. Ng}.}
  \bibinfo{year}{2013}\natexlab{}.
\newblock \showarticletitle{Topic Segmentation and Labeling in Asynchronous
  Conversations}.
\newblock \bibinfo{journal}{{\em J. Artif. Int. Res.\/}} \bibinfo{volume}{47},
  \bibinfo{number}{1} (\bibinfo{date}{May} \bibinfo{year}{2013}),
  \bibinfo{pages}{521--573}.
\newblock
\showISSN{1076-9757}
\showURL{%
\url{http://dl.acm.org/citation.cfm?id=2566972.2566986}}


\bibitem[\protect\citeauthoryear{Kim, Wang, and Baldwin}{Kim
  et~al\mbox{.}}{2010}]%
        {kim-tagging-2010}
\bibfield{author}{\bibinfo{person}{Su~Nam Kim}, \bibinfo{person}{Li Wang},
  {and} \bibinfo{person}{Timothy Baldwin}.} \bibinfo{year}{2010}\natexlab{}.
\newblock \showarticletitle{Tagging and linking web forum posts}. In
  \bibinfo{booktitle}{{\em CoNLL-2010}}. \bibinfo{pages}{192--202}.
\newblock


\bibitem[\protect\citeauthoryear{Kingma and Ba}{Kingma and Ba}{2014}]%
        {KingmaB14}
\bibfield{author}{\bibinfo{person}{Diederik~P. Kingma} {and}
  \bibinfo{person}{Jimmy Ba}.} \bibinfo{year}{2014}\natexlab{}.
\newblock \showarticletitle{Adam: {A} Method for Stochastic Optimization}.
\newblock \bibinfo{journal}{{\em CoRR\/}}  \bibinfo{volume}{abs/1412.6980}
  (\bibinfo{year}{2014}).
\newblock
\showURL{%
\url{http://arxiv.org/abs/1412.6980}}


\bibitem[\protect\citeauthoryear{Louis and Cohen}{Louis and Cohen}{2015}]%
        {louis2015conversation}
\bibfield{author}{\bibinfo{person}{Annie~P Louis} {and} \bibinfo{person}{Shay~B
  Cohen}.} \bibinfo{year}{2015}\natexlab{}.
\newblock \showarticletitle{Conversation trees: A grammar model for topic
  structure in forums}. Association for Computational Linguistics.
\newblock


\bibitem[\protect\citeauthoryear{McIntyre and Lapata}{McIntyre and
  Lapata}{2010}]%
        {McIntyre:2010}
\bibfield{author}{\bibinfo{person}{Neil McIntyre} {and}
  \bibinfo{person}{Mirella Lapata}.} \bibinfo{year}{2010}\natexlab{}.
\newblock \showarticletitle{Plot Induction and Evolutionary Search for Story
  Generation}. In \bibinfo{booktitle}{{\em Proceedings of the 48th Annual
  Meeting of the Association for Computational Linguistics}} {\em
  (\bibinfo{series}{ACL '10})}. \bibinfo{publisher}{Association for
  Computational Linguistics}, \bibinfo{address}{Uppsala, Sweden},
  \bibinfo{pages}{1562--1572}.
\newblock


\bibitem[\protect\citeauthoryear{Nguyen and Joty}{Nguyen and Joty}{2017}]%
        {dat-joty:2017}
\bibfield{author}{\bibinfo{person}{Dat~Tien Nguyen} {and}
  \bibinfo{person}{Shafiq Joty}.} \bibinfo{year}{2017}\natexlab{}.
\newblock \showarticletitle{A Neural Local Coherence Model}. In
  \bibinfo{booktitle}{{\em Proceedings of the 55th Annual Meeting of the
  Association for Computational Linguistics (Volume 1: Long Papers)}} {\em
  (\bibinfo{series}{ACL '17})}. \bibinfo{publisher}{Association for
  Computational Linguistics}, \bibinfo{address}{Vancouver, Canada},
  \bibinfo{pages}{(to appear)}.
\newblock


\bibitem[\protect\citeauthoryear{Pitler, Louis, and Nenkova}{Pitler
  et~al\mbox{.}}{2010}]%
        {Pitler:2010}
\bibfield{author}{\bibinfo{person}{Emily Pitler}, \bibinfo{person}{Annie
  Louis}, {and} \bibinfo{person}{Ani Nenkova}.}
  \bibinfo{year}{2010}\natexlab{}.
\newblock \showarticletitle{Automatic Evaluation of Linguistic Quality in
  Multi-document Summarization}. In \bibinfo{booktitle}{{\em Proceedings of the
  48th Annual Meeting of the Association for Computational Linguistics}} {\em
  (\bibinfo{series}{ACL '10})}. \bibinfo{publisher}{Association for
  Computational Linguistics}, \bibinfo{address}{Uppsala, Sweden},
  \bibinfo{pages}{544--554}.
\newblock


\bibitem[\protect\citeauthoryear{Srivastava, Hinton, Krizhevsky, Sutskever, and
  Salakhutdinov}{Srivastava et~al\mbox{.}}{2014}]%
        {Srivastava14a}
\bibfield{author}{\bibinfo{person}{Nitish Srivastava},
  \bibinfo{person}{Geoffrey Hinton}, \bibinfo{person}{Alex Krizhevsky},
  \bibinfo{person}{Ilya Sutskever}, {and} \bibinfo{person}{Ruslan
  Salakhutdinov}.} \bibinfo{year}{2014}\natexlab{}.
\newblock \showarticletitle{Dropout: A Simple Way to Prevent Neural Networks
  from Overfitting}.
\newblock \bibinfo{journal}{{\em Journal of Machine Learning Research\/}}
  \bibinfo{volume}{15} (\bibinfo{year}{2014}), \bibinfo{pages}{1929--1958}.
\newblock


\bibitem[\protect\citeauthoryear{Tieleman and Hinton}{Tieleman and
  Hinton}{2012}]%
        {Tieleman12}
\bibfield{author}{\bibinfo{person}{T. Tieleman} {and} \bibinfo{person}{G
  Hinton}.} \bibinfo{year}{2012}\natexlab{}.
\newblock \showarticletitle{{RMSprop}}.
\newblock  (\bibinfo{year}{2012}).
\newblock


\bibitem[\protect\citeauthoryear{Wang, Lui, Kim, Nivre, and Baldwin}{Wang
  et~al\mbox{.}}{2011}]%
        {Wang:2011:PTD}
\bibfield{author}{\bibinfo{person}{Li Wang}, \bibinfo{person}{Marco Lui},
  \bibinfo{person}{Su~Nam Kim}, \bibinfo{person}{Joakim Nivre}, {and}
  \bibinfo{person}{Timothy Baldwin}.} \bibinfo{year}{2011}\natexlab{}.
\newblock \showarticletitle{Predicting Thread Discourse Structure over
  Technical Web Forums}. In \bibinfo{booktitle}{{\em Proceedings of the
  Conference on Empirical Methods in Natural Language Processing}} {\em
  (\bibinfo{series}{EMNLP '11})}. \bibinfo{publisher}{Association for
  Computational Linguistics}, \bibinfo{address}{Stroudsburg, PA, USA},
  \bibinfo{pages}{13--25}.
\newblock
\showISBNx{978-1-937284-11-4}
\showURL{%
\url{http://dl.acm.org/citation.cfm?id=2145432.2145435}}


\bibitem[\protect\citeauthoryear{Wang, Joshi, Cohen, and Ros{\'e}}{Wang
  et~al\mbox{.}}{2008a}]%
        {wang-recovering-2008}
\bibfield{author}{\bibinfo{person}{Yi-Chia Wang}, \bibinfo{person}{Mahesh
  Joshi}, \bibinfo{person}{William Cohen}, {and} \bibinfo{person}{Carolyn
  Ros{\'e}}.} \bibinfo{year}{2008}\natexlab{a}.
\newblock \showarticletitle{Recovering implicit thread structure in newsgroup
  style conversations}. In \bibinfo{booktitle}{{\em AAAI}}.
\newblock


\bibitem[\protect\citeauthoryear{Wang, Joshi, Cohen, and Rosé}{Wang
  et~al\mbox{.}}{2008b}]%
        {yi:2008}
\bibfield{author}{\bibinfo{person}{Yi-Chia Wang}, \bibinfo{person}{Mahesh
  Joshi}, \bibinfo{person}{William Cohen}, {and} \bibinfo{person}{Carolyn
  Rosé}.} \bibinfo{year}{2008}\natexlab{b}.
\newblock \showarticletitle{Recovering Implicit Thread Structure in Newsgroup
  Style Conversations}. In \bibinfo{booktitle}{{\em Proceedings of the Eleventh
  International Conference on Web and Social Media, {ICWSM} 2008.}}
\newblock


\bibitem[\protect\citeauthoryear{Zeiler}{Zeiler}{2012}]%
        {Zeiler12}
\bibfield{author}{\bibinfo{person}{Matthew~D. Zeiler}.}
  \bibinfo{year}{2012}\natexlab{}.
\newblock \showarticletitle{{ADADELTA:} An Adaptive Learning Rate Method}.
\newblock \bibinfo{journal}{{\em CoRR\/}}  \bibinfo{volume}{abs/1212.5701}
  (\bibinfo{year}{2012}).
\newblock
\showURL{%
\url{http://arxiv.org/abs/1212.5701}}


\end{thebibliography}

\end{document}